\documentclass[12pt]{article}
\usepackage{amsmath}
\usepackage{amssymb}
\usepackage{graphicx}
\usepackage{setspace}
\usepackage{graphicx}
\usepackage{epsfig}
\usepackage{amsfonts}
\usepackage{dsfont}
\usepackage{bm}
\usepackage{color}
\usepackage{comment}
\usepackage[dvipsnames]{xcolor}
\usepackage[authoryear,round]{natbib}
\usepackage{url}
\usepackage{pgfplots}
\usepackage[hidelinks]{hyperref}
\hypersetup{
  colorlinks   = true, %Colours links instead of ugly boxes
  urlcolor     = blue, %Colour for external hyperlinks
  linkcolor    = blue, %Colour of internal links
  citecolor   = blue %sepia %Colour of citations
}
\usepackage[T1]{fontenc}
\usepackage[cal=boondoxo]{mathalfa}

\usepackage{soul}

\newtheorem{corollary}{\bf Corollary}

\newtheorem{proposition}{\bf Proposition}

\newtheorem{lemma}{\bf Lemma}

\newtheorem{example}{\bf Example}
\newenvironment{proof}[1][Proof]{\noindent\textbf{#1} }{\ \rule{0.5em}{0.5em}}

\renewcommand{\ge}{\geqslant}
\renewcommand{\le}{\leqslant}

\pgfplotsset{compat=1.17}
\DeclareMathOperator{\dd}{\textrm{d}\!}
\allowdisplaybreaks

\begin{document}
\title{The psychology of prizes:\\ Loss aversion and optimal tournament rewards}
\author{Dmitry Ryvkin\thanks{Economics Discipline Group, School of Economics, Finance and Marketing, RMIT University,  \url{d.ryvkin@gmail.com}.} \and Qin Wu\thanks{Economics Discipline Group, School of Economics, Finance and Marketing, RMIT University,  \url{qin.wu@rmit.edu.au}.}}
\date{This version: \today}

\maketitle

\begin{abstract}
\noindent We study the optimal allocation of prizes in rank-order tournaments with loss averse agents. Prize sharing becomes increasingly optimal with loss aversion because more equitable prizes reduce the marginal psychological cost of anticipated losses. Furthermore, loss aversion can boost effort if prizes are sufficiently equitable, but otherwise effort declines with loss aversion. Overall, these results give credence to more equitable allocations of competitive rewards. A win-win scenario is where optimal prizes are equitable even under loss neutrality, in which case the principal benefits from agents' loss aversion.

\bigskip


\end{abstract}

\newpage
\onehalfspacing

\section{Introduction}

In this paper, we study the optimal allocation of competitive prizes, such as rank-based bonuses, in a setting where agents are \emph{loss averse}. That is, in addition to standard monetary payoffs, agents derive psychological utility from anticipated gains and losses relative to an expectations-based reference point \citep{kHoszegi-Rabin:2006}, where the marginal impact of losses is stronger relative to gains. We use a rank-order tournament model similar to \cite{Lazear-Rosen:1981}, with stochastic performance affected by costly effort as well as idiosyncratic additive noise, where a principal allocates a fixed budget to rank-dependent prizes. We study the impact of loss aversion on the agents' effort and on the optimal (effort-maximizing) prize schedule.

Rank-based rewards are common in organizations and other settings. For example, end-of-year bonuses or salary raises are typically based on performance rankings, with managers allocating a fixed budget to these incentives. Therefore, one of the central questions organizations face is that of \emph{optimal prize allocation}: How should these rank-based rewards be structured to maximize workers' performance? This question has been addressed in the literature using various models of rank-based rewards, referred to as contests or tournaments, where agents have standard preferences. It is well-known, however, that most individuals' preferences exhibit reference dependence and loss aversion \citep{Kahneman-Tversky:1979}, whereby outcomes are evaluated as gains or losses relative to a reference point, and the marginal disutility of a loss exceeds the marginal utility of a gain by a factor $\lambda>1$ referred to as the \emph{loss aversion coefficient}. According to a recent meta-analysis by \cite{Brown-et-al:2024}, the average value of $\lambda$ across more than a hundred experimental studies is close to 2, indicating substantial loss aversion in the population.

The impact of loss aversion on performance has been studied in various domains. In sports, loss aversion forces individuals and team managers to use safer, and ultimately suboptimal, strategies when they are ahead \citep{Pope-Schweitzer:2011,Pedace-Smith:2013}. Similarly, in education, loss-averse students tend to perform worse on multiple-choice exams where unanswered questions receive partial credit, as they avoid guessing to secure guaranteed points \citep{Karle-et-al:2022}. Managerial loss aversion has been linked to inferior investment decisions \citep{Odean:1998,Fellner-Sutter:2009}, suboptimal operational decisions \citep{schweitzer2000decision,wang2009would}, and even an increased incidence of illegal behavior \citep{Mishina-et-al:2010}. Thus, it is generally believed that loss aversion is detrimental to performance, and individuals need to be ``debiased,'' i.e., trained to be less loss averse, to realize their full potential \citep{Hueber-Schwaiger:2022}.

At the same time, it has been demonstrated how loss aversion can be harnessed to generate extra incentives for effort provision. In a field experiment with factory workers, \cite{Hossain-List:2012} find that framing compensation as a punishment leads to higher effort than when it is framed as a reward. \cite{Fryer-et-al:2012} report similar results for the impact of teacher bonuses on student achievement; and \cite{Abeler-et-al:2011} show that by manipulating reference points (expectations of earnings) it is possible to make individuals work longer. 

These findings suggest that the impact of loss aversion depends on the incentive scheme, and performance can be improved, or at least the negative impact of loss aversion reduced, if incentives are properly adjusted. For example, \cite{Marchegiani-et-al:2016} show that it helps for managers to be lenient when workers are loss averse; and \cite{Dodonova-Khoroshilov:2006} show that compensation packages with a large share of stock options are optimal for loss averse managers. In this paper, we further contribute to understanding how loss aversion interacts with incentives, and how it can be harnessed to improve performance.

In a tournament setting, agents exert effort, and outcomes are realized in the form of prizes. Following \cite{kHoszegi-Rabin:2006,kHoszegi-Rabin:2007}, we posit that any given prize is evaluated relative to a stochastic reference point represented by all alternative outcomes, i.e., all other possible prizes the agent could receive. This reference point is endogenous, in the sense that the probabilities of alternative prizes are influenced by the agent's effort decision as well as the efforts of other agents. We adopt the \emph{choice-acclimating personal equilibrium} solution concept, whereby agents correctly anticipate how their efforts will affect those probabilities.

We have two major results. First, the equilibrium effort can decrease or increase with loss aversion, depending on the allocation of prizes. Specifically, for any distribution of noise, effort decreases with loss aversion when fewer than half of all agents receive a positive prize; and increases with loss aversion when more than half of all agents receive the same top prize. For more general prize allocations, the results are more nuanced and depend also on the details of the noise distribution.

Second, for any distribution of noise, the optimal allocation of prizes becomes (weakly) more equitable as loss aversion rises. We know from the existing literature that the exact configuration of optimal prizes in rank-order tournaments (without loss aversion) is determined by the distribution of noise. Specifically, optimal prizes tend to be more equitable for noise distributions with heavier tails \citep{Drugov-Ryvkin:2020_prizes_JET}. While those forces continue to operate also in the presence of loss aversion, the latter provides an additional mechanism for the optimality of prize sharing. 

Combining the two results, we find that for distributions of noise such that relatively unequal prizes are optimal for loss neutral agents, the equilibrium effort can be nonmonotone (specifically, U-shaped) in loss aversion if prizes are adjusted optimally. The reason is that for relatively low loss aversion the optimal number of positive prizes is small and effort declines, whereas for sufficiently high loss aversion the optimal number of equal prizes at the top becomes large enough and effort then rises.   

For intuition, recall that each prize is evaluated relative to all other prizes which can be received with the corresponding probabilities; therefore, the contribution of each prize $v_r$ awarded for rank $r$ to the gain-loss utility is a linear combination of differences $v_r-v_s$, for all $s\ne r$, with weights equal to the product of probabilities of receiving $v_r$ and $v_s$. In the symmetric equilibrium, all these probabilities are equal, but, due to loss aversion, losses (i.e., negative differences $v_r-v_s$) contribute more to the psychological utility as compared to gains, and hence the overall psychological utility is always negative. 

The psychological \emph{marginal} utility of effort, however, can be positive or negative, and its sign depends on both the distribution of noise and the allocation of prizes. Similar to the psychological utility itself, the marginal psychological utility is controlled by \emph{loss states}, i.e., by events with negative differences $v_r-v_s$ for each prize $v_r$. The marginal probability of such an event is given by the sum of marginal probabilities of receiving prizes $v_r$ and $v_s$, which is (typically) positive when these prizes are close to the top, negative when they are close to the bottom, and has an ambiguous sign for ranks in the middle. The reason is that an increase in effort raises the probability of reaching a high rank and lowers the probability of reaching a low rank. Therefore, if prize differences at the top are relatively small, for example, when there are many equal prizes awarded at the top, the overall marginal probability of the psychological loss tends to be negative, and hence the equilibrium effort rises with loss aversion. The opposite holds when there are few equal prizes at the top, i.e., when the prize allocation is less equitable. 

Broadly, our results imply that prize sharing is beneficial when agents are loss averse, and loss aversion can boost productivity if prizes are sufficiently equitable. In environments with light-tailed shocks, highly unequal prize schedules are optimal under loss neutrality; therefore, agents' loss aversion is detrimental for the organization. The negative impact of loss aversion on effort can be partially mitigated by making prizes more equitable, but the organization would still benefit from hiring less loss averse workers, or training its workers to be less loss averse. In contrast, in environments with heavy-tailed shocks equitable prize schedules are optimal even for loss neutral agents. In this case, somewhat surprisingly, agents' loss aversion helps the organization, and optimal prize sharing helps boost productivity further when loss aversion rises. Hiring more loss averse workers would benefit the organization in such a setting. 

Heavy-tailed shocks are characteristic of new and post-industrial sectors, such as sales of creative or innovative products \citep{DeVany-Walls:2004}, R\&D \citep{Casault-et-al:2017}, asset prices \citep{Rachev:2003}, or returns to venture capital investment \citep{deTreville-et-al:2014}. Our results then suggest that a combination of equitable prizes and loss averse agents could benefit firms in these sectors. One implication is that hiring more women, who are typically under-represented in these sectors and tend to be relatively more loss averse \citep{VonGaudecker-et-al:2011,Andersson-et-al:2016,Dawson:2023}, could boost productivity if prizes are optimally adjusted.

\paragraph{Related literature} Our paper combines a prominent model of reference dependence and loss aversion with endogenous, expectation-based reference points \citep{kHoszegi-Rabin:2006,kHoszegi-Rabin:2007} with a classical model of rank-order tournaments with stochastic performance \citep{Lazear-Rosen:1981}. While several authors studied tournaments and contests with loss averse players,\footnote{Relatedly, \cite{Eisenhuth-Grunewald:2020} and \cite{Balzer-Rosato:2021} consider auctions with loss averse bidders.} none of them addressed the optimal prize allocation problem.

\cite{Gill-Stone:2010} explore two-player tournaments with \emph{desert} where agents experience gains and losses relative to a reference point calculated as their expected prize share in the tournament. \cite{Fallucchi-Trevisan:2024} study a similar model for $n$-player winner-take-all Tullock contests. \cite{Lim:2010} considers a social reference point based on the modal payoff in the group, akin to inequity aversion \`{a} la \cite{Fehr-Schmidt:1999}. 

Our setting is most similar to that of \cite{Dato-et-al:2018} and \cite{Fu-et-al:2022} who consider loss aversion with expectation-based reference points following \cite{kHoszegi-Rabin:2006,kHoszegi-Rabin:2007}. \cite{Dato-et-al:2018} study
two-player tournaments and compare the implications of different solution concepts: choice-acclimating personal equilibrium, preferred personal equilibrium and personal equilibrium; whereas \cite{Fu-et-al:2022} focus on the choice-acclimating personal equilibrium (as we do in this paper) and study generalized winner-take-all Tullock contests.

%The paper most closely related to ours is by \cite{Lim:2010} who studies how loss aversion affects effort in rank-order tournaments with different numbers of winners (i.e., equal, positive prizes) and shows that if loss aversion is sufficiently strong then effort increases with the number of winner prizes. While this finding is similar in spirit to our Proposition \ref{prop_effort_kappa}, we note that \cite{Lim:2010} assumes the reference point is exogenous and based on the modal prize; thus, \emph{social} loss aversion in his model is akin to inequity aversion \`{a} la \cite{Fehr-Schmidt:1999}. Therefore, prize inequality interacts with loss aversion through a different---social comparison---mechanism in his model, and more equitable prize allocations make these comparisons more favorable, on average. In contrast, in our model loss aversion operates through expectations of alternative outcomes, with endogenous probabilities. Additionally, \cite{Lim:2010} does not study the optimal prize allocation problem; and he only considers the uniform distribution of noise,\footnote{We show that the winner-take-all prize schedule is optimal when noise is uniformly distributed for any loss aversion, despite the fact that effort decreases with loss aversion in this case.} and hence there is no exploration of how the effect of loss aversion depends on the shape of the distribution of noise.

Finally, our paper is related to the literature on the optimal allocation of prizes in tournaments and contests without loss aversion. \cite{Moldovanu-Sela:2001}, \cite{Fang-et-al:2020}, and \cite{Drugov-Ryvkin:2020_prizes_JET} study this question for all-pay contests under incomplete information, all-pay contests under complete information, and rank-order tournaments with stochastic performance, respectively. In particular, most of the machinery we use in this paper to deal with comparative statics related to the properties of noise is developed by \cite{Drugov-Ryvkin:2020_prizes_JET}.

\bigskip

The rest of the paper is structured as follows. Section \ref{sec:model} presents the model. The equilibrium is characterized in Section \ref{sec:equilibrium}. Our main results on the effect of loss aversion on effort and the optimal allocation of prizes are collected in Sections \ref{sec:effort} and \ref{sec:prizes}, respectively. Section \ref{sec:conclusions} concludes. 

\section{The model}
\label{sec:model}

There are $n\ge 2$ identical agents, indexed by $i=1,\ldots,n$, who participate in a tournament by simultaneously and independently exerting efforts $x_i\in\mathds{R}_+$. The cost of effort $x_i$ to agent $i$ is $c(x_i)$, where $c(\cdot)$ is a strictly increasing function such that $\bar{x}=c^{-1}(1)$ is finite. We assume that $c(\cdot)$ is strictly convex in $[0,\bar{x}]$, with $c(0)=c'(0)=0$. Each agent's output is stochastic and takes the form $Y_i=x_i+\epsilon_i$, where $\epsilon_i$ are random shocks drawn independently across agents from a common distribution $F(\cdot)$ with density $f(\cdot)$. 

The tournament pay scheme is a vector of prizes ${\bf v}=(v_1,\ldots,v_n)$ that are non-negative, decreasing, and sum up to one. Let $\mathcal{V}=\{{\bf v}\in\mathds{R}^n_+:v_1\ge\ldots\ge v_n\ge 0,\, \sum_{r=1}^nv_r=1\}$ denote the set of all such prize vectors. Prizes are assigned on the basis of output ranking; that is, the agent with the highest output earns $v_1$, the agent with the second highest output earns $v_2$, etc. Ties are a zero probability event in equilibrium.

The agents are loss averse, with identical preferences. We adopt a model of loss aversion with endogenous expectations-based reference points \`{a} la \cite{kHoszegi-Rabin:2006,kHoszegi-Rabin:2007}. As in \cite{Fu-et-al:2022}, the expectations are assumed to be \emph{choice-acclimating}, in the sense that the agents correctly anticipate how their actions (efforts) affect the probabilities of future outcomes (prizes) and base their expectations on those probabilities.

Formally, consider a strategy profile ${\bf x}=(x_1,\ldots,x_n)$ and let $p^{(i,r)}({\bf x})$ denote the probability for player $i$'s output to be ranked $r$. Player $i$'s utility then takes the form
\begin{align}
\label{utility}
u_i({\bf x}) = \sum_{r=1}^np^{(i,r)}({\bf x})v_r - c(x_i) + \eta\sum_{r=1}^np^{(i,r)}({\bf x})\sum_{s\ne r}p^{(i,s)}({\bf x})\mu(v_r-v_s).
\end{align}
The first two terms represent the standard expected monetary benefit and cost of effort, and the third term is the psychological utility of anticipated gains and losses. The cost of effort is incurred upfront, with certainty, and does not depend on the outcome of the tournament; therefore, it does not enter the evaluation of expected gains and losses. Parameter $\eta>0$ is the relative strength of the psychological component of utility. The latter is a sum over all mutually exclusive outcomes, with probabilities $p^{(i,r)}({\bf x})$, where for each outcome the agent's expected gain-loss from the realized prize, $v_r$, is evaluated against all possible alternative prizes, $v_s$, $s\ne r$, forming a stochastic reference point.\footnote{Thus, our equilibrium concept is \emph{choice-acclimating personal equilibrium} of \cite{kHoszegi-Rabin:2007} where actions (efforts) generate probabilities of future outcomes (prizes) and reference points (alternative prizes) that agents correctly anticipate. This assumption is justified if there is a sufficient time lag between effort choices and the realization of the prizes, which is reasonable for bonuses in organizations \citep{Fu-et-al:2022}. The same utility function would arise in a model of disappointment aversion \citep{Loomes-Sugden:1986}.} Function $\mu(v) = v + (\lambda-1)v\mathds{1}_{v<0}$ is the standard piece-wise linear psychological gain-loss utility, with loss aversion coefficient $\lambda\ge 1$.

It is straightforward to check that $\sum_{r=1}^np^{(i,r)}({\bf x})\sum_{s\ne r}p^{(i,s)}({\bf x})(v_r-v_s)=0$; in other words, in the absence of loss aversion the anticipated gains and losses cancel each other out. Using the structure of function $\mu(\cdot)$, we, therefore, rewrite (\ref{utility}) as
\begin{align}
\label{utility_loss}
u_i({\bf x}) = \sum_{r=1}^np^{(i,r)}({\bf x})v_r - c(x_i) - \theta\sum_{r=1}^np^{(i,r)}({\bf x})\sum_{s<r}p^{(i,s)}({\bf x})(v_s-v_r),
\end{align}
where $\theta:=\eta(\lambda-1)$. Note that the psychological component of utility is negative and arises from anticipated psychological losses associated with higher alternative prizes. In what follows, we assume $\theta \le 1$, i.e., this psychological utility is not too large as compared to the monetary utility.

\section{The equilibrium}
\label{sec:equilibrium}

We look for a symmetric, pure strategy equilibrium in which all agents exert the same effort $x^*>0$. Let $p^{(r)}(x;x^*)$ denote the probability for an agent with effort $x$ to be ranked $r$ given that all other agents' effort is $x^*$. From (\ref{utility_loss}), the indicative agent's utility is
\begin{align}
\label{utility_symm}
u(x;x^*) = \sum_{r=1}^np^{(r)}(x;x^*)v_r - \theta\sum_{r=1}^np^{(r)}(x;x^*)\sum_{s<r}p^{(s)}(x;x^*)(v_s-v_r) - c(x).
\end{align}
Probability $p^{(r)}(x;x^*)$ is given by
\begin{align}
\label{p_r}
p^{(r)}(x;x^*) = \binom{n-1}{r-1}\int F(x-x^*+t)^{n-r}[1-F(x-x^*+t)]^{r-1}\dd F(t).
\end{align}
Indeed, in order to be ranked $r$, the indicative agent's output must be above the output of $n-r$ other agents and below the output of $r-1$ other agents; and the binomial coefficient accounts for the number of equivalent ways the identities of these agents can be chosen. 

We adopt the first order approach, i.e., we look for a symmetric equilibrium as a solution to the symmetric first order condition $u_x(x^*;x^*)=0$. The equilibrium existence and uniqueness can be guaranteed by imposing assumptions on the curvature of the cost function $c(\cdot)$ and the dispersion of noise.\footnote{Specifically, we assume that $c_0=\inf_{x\in[0,\bar{x}]}c''(x)>0$ (this condition holds for a quadratic cost function $c(x)=\frac{c_0x^2}{2}$) and $|p^{(r)}_x|^2,|p^{(r)}_{xx}|$ are small compared to $c_0$. The latter can be achieved by considering a distribution of noise with a scale parameter, $F(\frac{t}{\sigma})$, for $\sigma$ large enough. Thus, we ensure that $u(x;x^*)$ is globally strictly concave in $x$. Finally, we assume that the participation constraint $u(x^*;x^*)\ge u_0$ is never binding, which is the case if the utility of not participating in the tournament, $u_0$, is sufficiently low. This is a reasonable assumption in an organizational setting where all employees are automatically evaluated and non-participation essentially means the worker resigns.} Of special importance are coefficients $\beta_r := p^{(r)}_x(x^*;x^*)$ given by
\begin{align}
\label{beta_r}
\beta_r = \binom{n-1}{r-1}\int F(t)^{n-r-1}[1-F(t)]^{r-2}[n-r - (n-1)F(t)]f(t)\dd F(t).
\end{align}
Each coefficient $\beta_r$ is the marginal effect of effort on the probability for the indicative agent to be ranked $r$ in the symmetric equilibrium. It is clear intuitively, and can be checked formally, that $\beta_1>0$, $\beta_n<0$, and $\sum_{r=1}^n\beta_r=0$. Indeed, by raising effort an agent always increases (respectively, decreases) her probability of being ranked first (respectively, last); and the total probability of getting one of the ranks does not change.

The symmetric first order condition takes the form
\begin{align}
\label{FOC}
& M({\bf v},\theta)=c'(x^*), \quad M({\bf v},\theta):=R({\bf v}) + \theta L({\bf v}),\\
& R({\bf v}) := \sum_{r=1}^n\beta_rv_r, \quad L({\bf v}) := - \frac{1}{n}\sum_{r=1}^n\sum_{s<r}[\beta_r(v_s-v_r)+\beta_s(v_s-v_r)].\nonumber
\end{align}
Here, $M({\bf v},\theta)$ is the equilibrium marginal benefit of effort that consists of two components: $R({\bf v})$ is the monetary marginal benefit; and $\theta L({\bf v})$ is the psychological marginal benefit of effort from anticipated gains and losses. Note that although the psychological utility term in (\ref{utility_symm}) is always negative, the psychological \emph{marginal} benefit can be positive or negative depending on the prize allocation and the distribution of noise (which determines coefficients $\beta_r$). This implies that (i) the equilibrium effort $x^*$ may go up or down as agents become more loss averse; and (ii) the optimal allocation of prizes may depend on loss aversion. In the next two sections we explore these effects in detail.

\section{How effort is affected by loss aversion}
\label{sec:effort}

Since $c(\cdot)$ is strictly convex, the right-hand side of (\ref{FOC}) is strictly increasing in $x^*$. It is then clear that the equilibrium effort $x^*$ increases (respectively, decreases) with loss aversion if $L({\bf v})$ is positive (respectively, negative).

Let $B_r:=\sum_{k=1}^r\beta_k$. Coefficient $B_r$ is a cumulative version of $\beta_r$ and, as such, represents the marginal effect of effort on the probability for the indicative agent to be ranked \emph{at least} $r$. It is then clear intuitively that $B_r>0$ for $r=1,\ldots,n-1$ and $B_n=0$. By transforming (\ref{beta_r}), it can be shown that
\begin{align}
\label{B_r}
B_r = r\binom{n-1}{r}\int F(t)^{n-1-r}[1-F(t)]^{r-1}f(t)\dd F(t).
\end{align}
Next, we introduce prize differentials $d_r:=v_r-v_{r+1}\ge 0$ for $r=1,\ldots,n-1$. The following lemma then provides a representation for $L({\bf v})$ that is instrumental for our results. All missing proofs are collected in Appendix \ref{app_proofs}.
\begin{lemma}
\label{lemma_L}
The psychological marginal benefit of anticipated gains and losses is $\theta L({\bf v})$, where
\begin{align}
\label{L_lemma} 
L({\bf v}) = \sum_{r=1}^{n-1}\left(\frac{2r}{n}-1\right)B_rd_r.
\end{align}
\end{lemma}
Representation (\ref{L_lemma}) shows that $L({\bf v})$ is a linear combination of (non-negative) marginal gains $B_rd_r$ from improving one's rank from $r+1$ to $r$. Such a marginal gain is positive only if the corresponding prize differential $d_r>0$. The weights in the linear combination (\ref{L_lemma}) are negative for $r<\frac{n}{2}$ and positive for $r>\frac{n}{2}$, indicating that prize differentials in the upper (respectively, lower) half of the rankings have a negative (respectively, positive) impact on the marginal gain-loss utility. This leads to the following proposition.

\begin{proposition}
\label{prop_effort_kappa}
For any prize schedule with at most $\lfloor\frac{n}{2}\rfloor$ positive prizes, the equilibrium effort decreases with loss aversion. For any prize schedule with at least $\lceil\frac{n}{2}\rceil$ equal prizes at the top, the equilibrium effort increases with loss aversion.
\end{proposition}

The following corollary considers a special class of prize schedules that divide the prize budget equally among $s$ top performers. Such prize schedules have only one positive prize differential $d_s>0$. They are special because, as we show in the next section, \emph{optimal} prize schedules generically have this form.

\begin{corollary}
\label{corr_effort_kappa}
Consider a prize schedule ${\bf v}^s=(\frac{1}{s},\ldots,\frac{1}{s},0,\ldots,0)$ that awards $s\le n-1$ equal prizes. If $s<\frac{n}{2}$, the equilibrium effort decreases with loss aversion; if $s>\frac{n}{2}$, the equilibrium effort increases with loss aversion.
\end{corollary}

For intuition, consider a schedule ${\bf v}^s$ with $s<\frac{n}{2}$. Then any rank at $s$ or higher is a \emph{gain state} producing a gain of $\frac{1}{s}$, while ranks below $s$ are \emph{loss states} where agents experience a loss of $\frac{1}{s}$, relative to the alternative outcome. By marginally raising effort, an agent raises the probability of reaching a gain state and reduces the probability of a loss state. However, in each of the gain states, the anticipated psychological gains are based on the probability of ending up in a loss state, and vice versa. Because there are fewer gain states in this case, the overall marginal effect of effort on the gain-loss utility is negative, and hence effort falls with loss aversion. The opposite holds for $s>\frac{n}{2}$. 

For more general prize schedules covered by Proposition \ref{prop_effort_kappa}, the same intuition applies because loss states outweigh gain states when all ranks in the bottom half receive zero prizes, and the opposite holds when all agents in the top half receive the same prize. For even more general prize schedules, the effect of loss aversion is ambiguous and depends also on the details of the distribution of noise.\footnote{For example, if prizes are equidistant, i.e., $v_r=\frac{2(n-r)}{n(n-1)}$ and $d_r=\frac{2}{n(n-1)}$, and $f$ is unimodal and symmetric, then $L({\bf v})=0$ and effort is unaffected by loss aversion.}

\section{How optimal prizes are affected by loss aversion}
\label{sec:prizes}

The equilibrium effort $x^*$ is maximized by a prize schedule ${\bf v}\in\mathcal{V}$ that maximizes the left-hand side of (\ref{FOC}). The optimal allocation of prizes, therefore, solves $\max_{\bf v\in\mathcal{V}}M({\bf v},\theta)$---a linear programming problem whose solution is characterized as follows.

\begin{proposition}
\label{prop_r_opt_theta}
(i) Prize schedule ${\bf v}^{r^*(\theta)}$ awarding $r^*(\theta)$ equal prizes at the top is optimal, with
\begin{align}
\label{r_opt}
r^*(\theta) \in \arg\max\{A_r(\theta):1\le r\le n-1\}, \quad \text{ where } A_r(\theta):=\left[1+\theta\left(\frac{2r}{n}-1\right)\right]\frac{B_r}{r}.
\end{align}

(ii) The optimal number of top prizes $r^*(\theta)$ (weakly) rises with loss aversion.
\end{proposition}

Proposition \ref{prop_r_opt_theta} is a very general result showing that, regardless of the shape of the distribution of noise, more prize sharing becomes optimal as loss aversion rises. Still, this proposition tells us little about what the optimal prize allocation looks like and, in particular, how effort under the optimal prizes depends on loss aversion. 

Let $r^*_0=r^*(0)$ denote the optimal number of prizes without loss aversion. As seen from (\ref{r_opt}), $r^*_0$ maximizes coefficients $\bar\beta_r:=\frac{B_r}{r}$ over $r$. As shown by \cite{Drugov-Ryvkin:2020_prizes_JET}, these coefficients can be written as $\bar\beta_r=\frac{1}{n}\mathds{E}[h(X_{(n-r:n)})]$, where $X_{(n-r:n)}$ is the $(n-r)^{\rm th}$ order statistic\footnote{We follow \cite{David-Nagaraja:2004} and enumerate order statistics so that $X_{(1:n)}\le\ldots\le X_{(n:n)}$. We also adopt the convention that $X_{(0:n)}=-\infty$ almost surely.} and $h(t)=\frac{f(t)}{1-F(t)}$ is the hazard (or failure) rate of noise. We use the standard abbreviations IFR (increasing failure rate) and DFR (decreasing failure rate) to designate distributions whose failure rate is, respectively, increasing and decreasing. Further, we call a density $f(\cdot)$ \emph{unimodal} if it is first increasing, then decreasing; and similarly say that a distribution has a \emph{unimodal failure rate} if it is first IFR, then DFR. All these notions are weak, i.e., distributions with monotone densities or monotone failure rates are special cases. It follows, in particular, that $\bar\beta_r$ is unimodal in $r$, with a mode $r^*_0$, when the distribution of noise has a unimodal failure rate.

Proposition \ref{prop_r_opt_theta} implies that $r^*(\theta)\ge r^*_0$, and the following result narrows it down further for the vast majority of noise distributions considered in practice.

\begin{proposition}
\label{prop_r_opt_location}
(i) Suppose $f$ is unimodal and has a unimodal failure rate. Then $r^*_0\le r^*(\theta)\le\hat{r}:=\max\{r:\beta_r>0\}$.

(ii) If $f$ is DFR then maximum prize sharing, $r^*(\theta)=n-1$, is optimal for any loss aversion.
\end{proposition}
\begin{example}[uniform distribution]
For $\epsilon_i$ distributed uniformly on $[-\frac{b}{2},\frac{b}{2}]$, we have $\beta_1=-\beta_n=\frac{1}{b}$ and $\beta_r=0$ for all $2\le r\le n-1$. Then $B_1=\ldots=B_{n-1}=\frac{1}{b}$, $\bar{\beta}_r = \frac{1}{br}$, and hence $A_r(\theta) = \frac{1-\theta}{br} + \frac{2\theta}{nb}$, which is decreasing in $r$ and is maximized by $r^*(\theta)=1$ for any $\theta\in[0,1]$. Thus, the winner-take-all prize schedule is optimal for any $\theta$ when noise is uniformly distributed. This is consistent with Proposition \ref{prop_r_opt_location} because $r^*_0=\hat{r}=1$ in this case.
\end{example}

In combination with Corollary \ref{corr_effort_kappa}, Proposition \ref{prop_r_opt_location} implies the following.
\begin{corollary}
\label{cor_r_location} Suppose $f$ is unimodal and has a unimodal failure rate. With prizes chosen optimally, 

(i) If $\hat{r}\le \lfloor\frac{n}{2}\rfloor$ then the equilibrium effort declines with loss aversion.

(ii) If $r^*_0>\lceil\frac{n}{2}\rceil$ then the equilibrium effort increases with loss aversion.

(iii) If $\hat{r}>\lceil\frac{n}{2}\rceil$ then the equilibrium effort increases with loss aversion for $\theta$ large enough.
\end{corollary}

\begin{figure}[tbp]
\centering
% Gumbel distribution

\begin{minipage}{0.3\textwidth}
     \begin{tikzpicture}[scale=0.55]
            \begin{axis}[
                xlabel={$\theta$},
                xlabel style={font=\Large}, 
                ylabel style={font=\Large}, 
                tick label style={font=\Large}, 
                domain=0:1,
                samples=100,
                axis lines=left,
                scaled y ticks=false,
                yticklabel style={/pgf/number format/fixed}, 
                tick align=outside,
                xmax=1.05,
                ymin=0, ymax=0.11,
                ytick={0.02, 0.04, 0.06, 0.08, 0.1}
            ]
                \addplot[color=black, thick, solid] 
                {(2/675) * (1 - (13 * x)/15) + (377 * (1 - (11 * x)/15))/66150 + (1174 * (1 - (3 * x)/5))/143325 + (1991 * (1 - (7 * x)/15))/191100 + (7793 * (1 - x/3))/630630 + (1959 * (1 - x/5))/140140};
                \addplot[color=black, thick, dashed] 
                {(2503 * (1 + x/5))/90090 + (835397 * (1 + (13 * x)/15))/270270000};
            \node[anchor=south] at (0.17,0.093) {\Large $M({\bf v};\theta)$};
            \node[anchor=south] at (1,0.016) {\Large ${\bf v}'$};
            \node[anchor=south] at (1,0.04) {\Large ${\bf v}''$};
            \end{axis}
        \end{tikzpicture}
\end{minipage}
\hfill
\begin{minipage}{0.3\textwidth}
\begin{tikzpicture}[scale=0.55]
        \begin{axis}[
            xlabel={$\theta$},
            xlabel style={font=\Large}, 
            ylabel style={font=\Large}, 
            tick label style={font=\Large}, 
            axis lines=left,
            xmax=1.05, 
            xmin=0,
            ymax=17,
            ymin=0,
            xtick={0, 0.2, 0.4, 0.6, 0.8, 1},
            ytick={5, 10, 15},
            tick align=outside 
        ]
            \addplot[const plot, thick, color=black] coordinates {
                (0, 1) (0.24, 1) (0.25, 2) (0.27, 3) (0.3, 4) (0.34, 5) (0.4, 6) (0.48, 7) (0.59, 8) (0.77, 9) (1., 9)
            };
        \node[anchor=south] at (0.13,14.5) {\Large $r^*(\theta)$};
        \end{axis}
    \end{tikzpicture}
\end{minipage}
\hfill
\begin{minipage}{0.3\textwidth}
 \begin{tikzpicture}[scale=0.55]
        \begin{axis}[
            xlabel={$\theta$},
            xlabel style={font=\Large}, 
            ylabel style={font=\Large}, 
            tick label style={font=\Large}, 
            samples=100,
            axis lines=left,
            scaled y ticks=false,
            yticklabel style={/pgf/number format/fixed}, 
            xmax=1.05,
            ymin=0, ymax=0.11,
            ytick={0.02, 0.04, 0.06, 0.08, 0.1},
            tick align=outside
        ]
            \addplot[mark=*, mark size=0.1pt, color=black] coordinates {
(0, 0.0622222) (0.01, 0.061683) (0.02, 0.0611437) (0.03, 0.0606044) 
(0.04, 0.0600652) (0.05, 0.0595259) (0.06, 0.0589867) (0.07, 0.0584474) 
(0.08, 0.0579081) (0.09, 0.0573689) (0.1, 0.0568296) (0.11, 0.0562904) 
(0.12, 0.0557511) (0.13, 0.0552119) (0.14, 0.0546726) (0.15, 0.0541333) 
(0.16, 0.0535941) (0.17, 0.0530548) (0.18, 0.0525156) (0.19, 0.0519763) 
(0.2, 0.051437) (0.21, 0.0508978) (0.22, 0.0503585) (0.23, 0.0498193) 
(0.24, 0.0493092) (0.25, 0.0488704) (0.26, 0.0484315) (0.27, 0.0480494) 
(0.28, 0.0477054) (0.29, 0.0473614) (0.3, 0.0470401) (0.31, 0.0467849) 
(0.32, 0.0465296) (0.33, 0.0462743) (0.34, 0.0460193) (0.35, 0.0458463) 
(0.36, 0.0456733) (0.37, 0.0455003) (0.38, 0.0453273) (0.39, 0.0451542) 
(0.4, 0.045012) (0.41, 0.0449141) (0.42, 0.0448163) (0.43, 0.0447184) 
(0.44, 0.0446206) (0.45, 0.0445227) (0.46, 0.0444249) (0.47, 0.044327) 
(0.48, 0.0442789) (0.49, 0.0442484) (0.5, 0.0442179) (0.51, 0.0441874) 
(0.52, 0.0441569) (0.53, 0.0441264) (0.54, 0.0440959) (0.55, 0.0440654) 
(0.56, 0.0440349) (0.57, 0.0440044) (0.58, 0.0439739) (0.59, 0.0439777) 
(0.6, 0.0440059) (0.61, 0.0440341) (0.62, 0.0440623) (0.63, 0.0440905) 
(0.64, 0.0441187) (0.65, 0.0441469) (0.66, 0.0441751) (0.67, 0.0442034) 
(0.68, 0.0442316) (0.69, 0.0442598) (0.7, 0.044288) (0.71, 0.0443162) 
(0.72, 0.0443444) (0.73, 0.0443726) (0.74, 0.0444008) (0.75, 0.044429) 
(0.76, 0.0444572) (0.77, 0.0445305) (0.78, 0.0446077) (0.79, 0.0446849) 
(0.8, 0.044762) (0.81, 0.0448392) (0.82, 0.0449164) (0.83, 0.0449936) 
(0.84, 0.0450707) (0.85, 0.0451479) (0.86, 0.0452251) (0.87, 0.0453023) 
(0.88, 0.0453794) (0.89, 0.0454566) (0.9, 0.0455338) (0.91, 0.045611) 
(0.92, 0.0456881) (0.93, 0.0457653) (0.94, 0.0458425) (0.95, 0.0459197) 
(0.96, 0.0459968) (0.97, 0.046074) (0.98, 0.0461512) (0.99, 0.0462284) 
(1, 0.0463055)
            };
        \node[anchor=south] at (0.15,0.093) {\Large $M^*(\theta)$};    
        \end{axis}
    \end{tikzpicture}
\end{minipage}

\vspace{0.5cm}

\begin{minipage}{0.3\textwidth}
% Pareto Distribution
\begin{tikzpicture}[scale=0.55]
            \begin{axis}[
                xlabel={$\theta$},
                xlabel style={font=\Large}, 
                ylabel style={font=\Large}, 
                tick label style={font=\Large}, 
                domain=0:1,
                samples=100,
                axis lines=left,
                scaled y ticks=false,
                yticklabel style={/pgf/number format/fixed}, 
                tick align=outside,
                xmax=1.05,
                ymin=0, ymax=0.11,
                xtick={0, 0.2, 0.4, 0.6, 0.8, 1},
                ytick={0.02, 0.04, 0.06, 0.08, 0.1}
            ]
                \addplot[color=blue, thick, solid] {(1 - (13 * x)/15)/2520 + 1/840 * (1 - (11 * x)/15) + 1/420 * (1 - (3 * x)/5) + 1/252 * (1 - (7 * x)/15) + 1/168 * (1 - x/3) + 1/120 * (1 - x/5)};
                \addplot[color=blue, thick, dashed] {3/100 *(1 + x/5) + 7/400* (1 + (13 * x)/15)};
                \node[anchor=south] at (0.17,0.093) {\Large $M({\bf v};\theta)$};
                \node[anchor=south] at (1,0.015) {\Large ${\bf v}'$};
                \node[anchor=south] at (1,0.055) {\Large ${\bf v}''$};
            \end{axis}
        \end{tikzpicture}

\end{minipage}
\hfill
\begin{minipage}{0.3\textwidth}
 \begin{tikzpicture}[scale=0.55]
        \begin{axis}[
            xlabel={$\theta$},
            xlabel style={font=\Large}, 
            ylabel style={font=\Large}, 
            tick label style={font=\Large}, 
            samples=100,
            axis lines=left,
            scaled y ticks=false,
            yticklabel style={/pgf/number format/fixed}, 
            xmax=1.05, 
            xmin=0,
            ymax=17,
            ymin=0,
            xtick={0, 0.2, 0.4, 0.6, 0.8, 1},
            ytick={5, 10, 15},
            tick align=outside
        ]
            \addplot [
                color=blue,
            ]
            {14};
        \node[anchor=south] at (0.13,14.5) {\Large $r^*(\theta)$};
        \end{axis}
    \end{tikzpicture}
\end{minipage}
\hfill
\begin{minipage}{0.3\textwidth}
\begin{tikzpicture}[scale=0.55]
        \begin{axis}[
            xlabel={$\theta$},
            xlabel style={font=\Large}, 
            ylabel style={font=\Large}, 
            tick label style={font=\Large}, 
            samples=100,
            axis lines=left,
            scaled y ticks=false,
            yticklabel style={/pgf/number format/fixed}, 
            xmax=1.05, 
            xmin=0,
            ymax=0.11,
            ymin=0,
            xtick={0, 0.2, 0.4, 0.6, 0.8, 1},
            ytick={0.02, 0.04, 0.06, 0.08, 0.1},
            tick align=outside
        ]
            \addplot [
                color=blue,
            ]
            {1/16 * (1 + (13 * x)/15)};
            \node[anchor=south] at (0.15,0.093) {\Large $M^*(\theta)$};
        \end{axis}
    \end{tikzpicture}
\end{minipage}

\vspace{0.5cm}

\begin{minipage}{0.3\textwidth}
% Burr distribution
        \begin{tikzpicture}[scale=0.55]
            \begin{axis}[
                xlabel={$\theta$},
                xlabel style={font=\Large}, 
                ylabel style={font=\Large}, 
                tick label style={font=\Large}, 
                domain=0:1,
                samples=100,
                axis lines=left,
                scaled y ticks=false,
                yticklabel style={/pgf/number format/fixed}, 
                tick align=outside,
                xmax=1.05,
                ymin=0, ymax=0.11,
                ytick={0.02, 0.04, 0.06, 0.08, 0.1}
            ]
                \addplot[color=red, thick, solid] {(334305 * pi * (1 - (13 * x)/15))/536870912 + (2414425 * pi * (1 - (11 * x)/15))/1610612736 + (676039 * pi * (1 - (3 * x)/5))/268435456 + (969969 * pi * (1 - (7 * x)/15))/268435456 + (2540395 * pi * (1 - x/3))/536870912 + (3128697 * pi * (1 - x/5))/536870912};
                \addplot[color=red, thick, dashed] {(96026931 * pi * (1 + x/5))/6710886400 + (13572783 * pi * (1 + (13 * x)/15))/5368709120};
                \node[anchor=south] at (0.17,0.093) {\Large $M({\bf v};\theta)$};
                \node[anchor=south] at (1,0.02) {\Large ${\bf v}'$};
                \node[anchor=south] at (1,0.055) {\Large ${\bf v}''$};
            \end{axis}
        \end{tikzpicture}
\end{minipage}
\hfill
\begin{minipage}{0.3\textwidth}
   \begin{tikzpicture}[scale=0.55]
        \begin{axis}[
            xlabel={$\theta$},
            xlabel style={font=\Large}, 
            ylabel style={font=\Large}, 
            tick label style={font=\Large}, 
            axis lines=left,
            xmax=1.05, 
            xmin=0,
            ymax=17,
            ymin=0,
            xtick={0, 0.2, 0.4, 0.6, 0.8, 1},
            ytick={5, 10, 15},
            tick align=outside 
        ]
            \addplot[const plot, thick, color=red] coordinates {
             (0, 7) (0.07, 8) (0.2, 9) (0.39, 10) (0.71, 11) (1,11)
            };
            \node[anchor=south] at (0.13,14.5) {\Large $r^*(\theta)$};
        \end{axis}
    \end{tikzpicture}
\end{minipage}
\hfill
\begin{minipage}{0.3\textwidth}
   \begin{tikzpicture}[scale=0.55]
        \begin{axis}[
            xlabel={$\theta$},
            xlabel style={font=\Large}, 
            ylabel style={font=\Large}, 
            tick label style={font=\Large}, 
            samples=100,
            axis lines=left,
            scaled y ticks=false,
            yticklabel style={/pgf/number format/fixed}, 
            xmax=1.05, 
            xmin=0,
            ymax=0.11, ymin=0,
            xtick={0, 0.2, 0.4, 0.6, 0.8, 1},
            ytick={0.02, 0.04, 0.06, 0.08, 0.1},
            tick align=outside
        ]
            \addplot[mark=*, mark size=0.1pt, color=red] coordinates {
(0, 0.0646169) (0.01, 0.0645738) (0.02, 0.0645307) (0.03, 0.0644876) 
(0.04, 0.0644445) (0.05, 0.0644015) (0.06, 0.0643584) (0.07, 0.0643774) 
(0.08, 0.0644201) (0.09, 0.0644629) (0.1, 0.0645056) (0.11, 0.0645483) 
(0.12, 0.064591) (0.13, 0.0646337) (0.14, 0.0646764) (0.15, 0.0647192) 
(0.16, 0.0647619) (0.17, 0.0648046) (0.18, 0.0648473) (0.19, 0.06489) 
(0.2, 0.0649328) (0.21, 0.0650576) (0.22, 0.0651825) (0.23, 0.0653074) 
(0.24, 0.0654322) (0.25, 0.0655571) (0.26, 0.065682) (0.27, 0.0658069) 
(0.28, 0.0659317) (0.29, 0.0660566) (0.3, 0.0661815) (0.31, 0.0663063) 
(0.32, 0.0664312) (0.33, 0.0665561) (0.34, 0.066681) (0.35, 0.0668058) 
(0.36, 0.0669307) (0.37, 0.0670556) (0.38, 0.0671804) (0.39, 0.067345) 
(0.4, 0.0675437) (0.41, 0.0677424) (0.42, 0.067941) (0.43, 0.0681397) 
(0.44, 0.0683383) (0.45, 0.068537) (0.46, 0.0687356) (0.47, 0.0689343) 
(0.48, 0.069133) (0.49, 0.0693316) (0.5, 0.0695303) (0.51, 0.0697289) 
(0.52, 0.0699276) (0.53, 0.0701262) (0.54, 0.0703249) (0.55, 0.0705236) 
(0.56, 0.0707222) (0.57, 0.0709209) (0.58, 0.0711195) (0.59, 0.0713182) 
(0.6, 0.0715169) (0.61, 0.0717155) (0.62, 0.0719142) (0.63, 0.0721128) 
(0.64, 0.0723115) (0.65, 0.0725101) (0.66, 0.0727088) (0.67, 0.0729075) 
(0.68, 0.0731061) (0.69, 0.0733048) (0.7, 0.0735034) (0.71, 0.0737338) 
(0.72, 0.0739923) (0.73, 0.0742507) (0.74, 0.0745092) (0.75, 0.0747676) 
(0.76, 0.0750261) (0.77, 0.0752845) (0.78, 0.075543) (0.79, 0.0758014) 
(0.8, 0.0760599) (0.81, 0.0763184) (0.82, 0.0765768) (0.83, 0.0768353) 
(0.84, 0.0770937) (0.85, 0.0773522) (0.86, 0.0776106) (0.87, 0.0778691) 
(0.88, 0.0781275) (0.89, 0.078386) (0.9, 0.0786445) (0.91, 0.0789029) 
(0.92, 0.0791614) (0.93, 0.0794198) (0.94, 0.0796783) (0.95, 0.0799367) 
(0.96, 0.0801952) (0.97, 0.0804537) (0.98, 0.0807121) (0.99, 0.0809706) 
(1, 0.081229)
            };
            \node[anchor=south] at (0.15,0.093) {\Large $M^*(\theta)$};
        \end{axis}
    \end{tikzpicture}
\end{minipage}

\caption{\textit{Left}: The equilibrium marginal benefit of effort, $M({\bf v},\theta)$, for prize schedules ${\bf v}'=(\frac{2}{7},\frac{5}{21},\frac{4}{21},\allowbreak \frac{1}{7},\frac{2}{21},\frac{1}{7},\allowbreak 0,0,0,0,\allowbreak 0,0,0,0,0)$ and ${\bf v}''=(\frac{1}{10},\frac{1}{10},\frac{1}{10},\allowbreak \frac{1}{10},\frac{1}{10},\frac{1}{10},\allowbreak \frac{1}{10},\frac{1}{10},\allowbreak \frac{1}{10},\frac{1}{50},\frac{1}{50},\frac{1}{50},\allowbreak \frac{1}{50},\frac{1}{50},0)$. \textit{Middle}: The optimal number of top prizes, $r^*(\theta)$. \textit{Right}: The equilibrium marginal benefit with optimal prizes, $M^*(\theta)$. \textit{Top row}: The Gumbel distribution with $F(t)=1-e^{-e^{-t}}$, ${\rm supp}(F)=\mathds{R}$, $B_r=(1-\frac{r}{n})(H_n-H_{n-r})$, where $H_k=\sum_{j=1}^k\frac{1}{j}$ is the $k$-th harmonic number. For $n=15$, $\hat{r}=8$ and $r^*_0=1$. \textit{Middle row}: The Pareto distribution with $F(t)=1-\frac{1}{t}$, ${\rm supp}(F)=[1,\infty)$, $B_r = \frac{r(r+1)}{n(n+1)}$, $r^*_0=\hat{r}=r^*(\theta)=n-1$. \textit{Bottom row}: The Burr distribution with $F(t)=1-\frac{1}{1+t^2}$, ${\rm supp}(F)=[0,\infty)$, $B_r = \frac{\pi(n-r)r(r+1)}{2^{2n-1}n(n+1)}\binom{2n-2r-1}{n-r}\binom{2r+1}{r}$. For $n=15$, $\hat{r}=11$ and $r^*_0=7$.}
\label{fig:examples}
\end{figure}

Propositions \ref{prop_effort_kappa} through \ref{prop_r_opt_location} are illustrated in Figure \ref{fig:examples} with examples considering tournaments with $n=15$ agents for three distributions of noise: Gumbel (top row), Pareto (middle row) and Burr (bottom row). The left panel in each row in Figure \ref{fig:examples} shows the dependence of the equilibrium marginal benefit of effort $M({\bf v},\theta$) on $\theta$ for two different prize schedules, ${\bf v}'$ and ${\bf v}''$. Schedule ${\bf v}'$ has 6 (different) positive prizes, and schedule ${\bf v}''$ has 9 equal prizes at the top. Consistent with Proposition \ref{prop_effort_kappa}, the equilibrium effort decreases with $\theta$ for the former and rises with $\theta$ for the latter.

The middle panel in each row in Figure \ref{fig:examples} shows the optimal number of top prizes, $r^*(\theta)$, as a function of $\theta$. Consistent with Proposition \ref{prop_r_opt_theta}, the optimal number of prizes weakly rises in all cases. The Gumbel distribution is IFR, with $r^*_0=1$ for any $n$ and $\hat{r}=8$ for $n=15$, and, consistent with Proposition \ref{prop_r_opt_location}, the optimal number of prizes changes between 1 and 8. The Pareto distribution is DFR, with $r^*_0=\hat{r}=n-1$ for all $n$; therefore, 14 prizes are optimal for any $\theta$. The Burr distribution has a unimodal failure rate, $r^*_0=7$, and $\hat{r}=11$; hence, the optimal number of prizes increases from 7 to 11 as $\theta$ increases.

Finally, the right panel in each row shows the dependence of the marginal benefit of effort with prizes optimally adjusted, $M^*(\theta):=M({\bf v}^{r^*(\theta)},\theta)$. The dependence of the equilibrium effort on $\theta$, of course, follows the same pattern. Effort rises with $\theta$ for $\theta$ large enough in all cases because $\hat{r}>\frac{n}{2}$ holds in all three examples. For the Gumbel distribution, it is still globally optimal to have loss neutral agents and the winner-take-all prize schedule. However, for the heavy-tailed Pareto and Burr distributions the highest effort is reached with sufficiently loss averse agents and equitable prizes.

\section{Conclusions}
\label{sec:conclusions}

We explored how prizes should be allocated in rank-order tournaments with loss averse agents. Loss aversion can affect effort positively or negatively, depending on the prize schedule. When prizes are too competitive, loss aversion is detrimental for effort; however, it boosts effort if prizes are sufficiently equitable. This result suggests that prize sharing can benefit the firm if its employees are loss averse. However, if prize sharing is not feasible, hiring less loss averse workers would be recommended.

Loss aversion also affects the optimal allocation of prizes, pushing it in the direction of more prize sharing when possible. In particular, in the presence of heavy-tailed shocks when prize sharing is optimal even without loss aversion, effort might rise with loss aversion if prizes are optimally adjusted. In such scenarios, the recommendation would be for the firm to hire more loss averse workers.

Our results contribute more broadly to understanding how behavioral biases interact with incentive schemes, and how the former can be harnessed to generate additional incentives and improve performance by carefully adjusting the latter.

\newpage
\bibliographystyle{aea}

\newpage
\appendix

\section{Proofs}
\label{app_proofs}

\begin{proof}{\bf of Lemma \ref{lemma_L}}
From (\ref{FOC}), we have
\begin{align*}
& L({\bf v}) = - \frac{1}{n}\sum_{r=1}^n\sum_{s<r}\left[\beta_r(v_s-v_r)+\beta_s(v_s-v_r)\right] \\
& = - \frac{1}{n}\left[\sum_{r=1}^n\sum_{s<r}\beta_rv_s - \sum_{r=1}^n\sum_{s<r}\beta_rv_r + \sum_{r=1}^n\sum_{s<r}\beta_sv_s - \sum_{r=1}^n\sum_{s<r}\beta_sv_r\right] \\
& = - \frac{1}{n}\left[\sum_{s=1}^{n-1}\sum_{r>s}\beta_rv_s - \sum_{r=1}^n(r-1)\beta_rv_r + \sum_{s=1}^{n-1}\sum_{r>s}\beta_sv_s - \sum_{r=1}^nB_{r-1}v_r\right]  \\
& = - \frac{1}{n}\left[\sum_{r=1}^{n-1}\sum_{s>r}\beta_sv_r - \sum_{r=1}^n(r-1)\beta_rv_r + \sum_{r=1}^{n-1}\sum_{s>r}\beta_rv_r - \sum_{r=1}^nB_{r-1}v_r\right] \\
& = - \frac{1}{n}\left[-\sum_{r=1}^{n-1}B_rv_r - \sum_{r=1}^n(r-1)\beta_rv_r + \sum_{r=1}^{n-1}(n-r)\beta_rv_r - \sum_{r=1}^nB_{r-1}v_r\right]\\
& = \frac{1}{n}\sum_{r=1}^n\left[2B_{r-1}-(n-2r)\beta_r\right]v_r.
\end{align*}
To obtain the fifth line, we used that $\sum_{s=1}^n\beta_s=0$ and hence $\sum_{s>r}\beta_s = -\sum_{s\le r}\beta_s = -B_r$. For the last line, we used that $B_{r-1}=B_r-\beta_r$.

We now use summation by parts to write $L({\bf v}) = \sum_{r=1}^n\ell_r(v_r-v_{r+1})$,\footnote{For convenience we define $v_{n+1}=0$, but ultimately it plays no role because $\ell_n=0$ as shown below.} where 
\begin{align*}
& \ell_r = \frac{1}{n}\sum_{k=1}^r\left[2B_{k-1}-(n-2k)\beta_k\right] = -B_r + \frac{2}{n}\sum_{k=1}^r(B_{k-1}+k\beta_k) \\
& = -B_r + \frac{2}{n}\left(\sum_{k=1}^r\sum_{s\le k-1}\beta_s + \sum_{k=1}^rk\beta_k\right) 
= -B_r + \frac{2}{n}\left(\sum_{s=1}^{r-1}\sum_{k=s+1}^r\beta_s + \sum_{k=1}^rk\beta_k\right) \\
& = -B_r + \frac{2}{n}\left(\sum_{s=1}^{r-1}(r-s)\beta_s + \sum_{k=1}^rk\beta_k\right) 
= -B_r + \frac{2}{n}\left(rB_{r-1}-\sum_{s=1}^{r-1}s\beta_s + \sum_{k=1}^rk\beta_k\right) \\
& = -B_r + \frac{2}{n}\left(rB_{r-1} +r\beta_r\right) = -B_r + \frac{2r}{n}B_r = \left(\frac{2r}{n}-1\right)B_r.
\end{align*}
Finally, recall that $B_n=0$, and hence the summation can stop at $r=n-1$.
\end{proof}

\bigskip

\begin{proof}{\bf of Proposition \ref{prop_effort_kappa}}
If there are at most $\lfloor\frac{n}{2}\rfloor$ positive prizes, we have $d_s=0$ for $s>\frac{n}{2}$. If there are at least $\lceil\frac{n}{2}\rceil$ equal prizes at the top, we have $d_s=0$ for $s<\frac{n}{2}$. The result then follows immediately from Lemma \ref{lemma_L}.
\end{proof}

\bigskip

\begin{proof}{\bf of Proposition \ref{prop_r_opt_theta}} Consider the problem 
\begin{align}
\label{problem_player_main}
\max[R({\bf v}) + \theta L({\bf v})] \quad \text{s.t. } \bf v\in\mathcal{V}.
\end{align}
{\bf Part (i)}: We first show that $v_n=0$, i.e., awarding zero prize to the bottom performer, is always optimal. From (\ref{FOC}) and (\ref{L_lemma}), the coefficient on $v_n$ in the objective is 
\[
\beta_n - \theta\left(\frac{2(n-1)}{n}-1\right)B_{n-1} = \left(1+\frac{\theta(n-2)}{n}\right)\beta_n<0,
\]
where we used that $B_{n-1} = B_n-\beta_n= -\beta_n$.

Further, using summation by parts we write $R({\bf v})=\sum_{r=1}^{n-1}B_rd_r$ and, with $v_n=0$, $\sum_{r=1}^nv_r = \sum_{r=1}^{n-1}rd_r$. Problem (\ref{problem_player_main}) then becomes
\begin{align}
\label{r_opt_problem}
\max \sum_{r=1}^{n-1}\left[1+\theta\left(\frac{2r}{n}-1\right)\right]B_rd_r \quad \text{s.t. } d_1,\ldots,d_{n-1}\ge 0, \, \sum_{r=1}^{n-1}rd_r=1.
\end{align}
Generically, this linear programming problem has a corner solution with $d_{r^*(\theta)}=\frac{1}{r^*(\theta)}$ and $d_r=0$ for $r\ne r^*(\theta)$, where $r^*(\theta)$ is given by (\ref{r_opt}).

\noindent {\bf Part (ii)}: Note that the maximization problem in (\ref{r_opt}) is equivalent to $\max_{r=1,\ldots,n-1}\log A_r(\theta)$, where
\[
\log A_r(\theta) = \log \left[1+\theta\left(\frac{2r}{n}-1\right)\right] + \log\frac{B_r}{r}.
\]
It is immediate that $\log A_r(\theta)$ has strictly increasing differences in $(r,\theta)$; and the result follows from the monotone comparative statics.
\end{proof}

\bigskip

\begin{proof}{\bf of Proposition \ref{prop_r_opt_location}}

\noindent {\bf Part (i)}: We rely on two results from \cite{Drugov-Ryvkin:2020_prizes_JET}: Lemma A.2 shows that if $f(\cdot)$ is unimodal then $B_r$ is unimodal in $r$; and Lemma 1 shows that if the noise has a unimodal failure rate then $\bar\beta_r$ is unimodal in $r$. It is clear that $\hat{r}$ is a mode of $B_r$, whereas $r^*_0$ is a mode of $\bar\beta_r$. Furthermore, $\hat{r}\ge r^*_0$. We can write the objective in (\ref{r_opt}) in the form
\[
A_r(\theta) = (1-\theta)\bar{\beta}_r + \frac{2\theta}{n}B_r,
\]
which is a convex combination of $\bar{\beta}_r$ and $\frac{2B_r}{n}$. Both are unimodal; therefore, the maximum of $A_r$ is reached at $r^*(\theta)$ between their respective maxima. Moreover, this maximum is at $r^*_0$ for $\theta=0$ and at $\hat{r}$ for $\theta=1$.

\noindent {\bf Part (ii)}: We know from the representation $\bar\beta_r=\frac{1}{n}\mathds{E}[h(X_{(n-r:n)})]$ that $\bar{\beta}_r$ is positive and increasing in $r$ (for $r=1,\ldots,n-1$) for DFR distributions. It is then clear that $A_r(\theta)$ is maximized by $r^*(\theta)=n-1$ for any $\theta\ge 0$.
\end{proof}

\end{document}